\documentclass{llncs}
\usepackage{amssymb,amsmath,subfigure,graphicx, pxfonts}
\usepackage{array}
\usepackage{rotating}
\usepackage{lmodern,blindtext}
\usepackage{epsfig}
\usepackage{multirow}
\usepackage{qtree}
\usepackage{stmaryrd}
\usepackage{ctable}

\begin{document}
\title{ A Rational and Efficient Algorithm for View Revision in Databases \thanks{This work extends from Chanderbose's [7].} }
\author{Radhakrishnan Delhibabu \thanks{The author acknowledges the support of RWTH Aachen, where
he is visiting scholar with an Erasmus Mundus External Cooperation
Window India4EU by the European Commission when the paper was
written.} \inst{1,2}\and Gerhard Lakemeyer \inst{2}}
\institute{Department of Computer Science and Engineering\\ SSN College of Engineering in Chennai, India\\
\email{delhibabur@ssn.edu.in}
\and Informatik 5, Knowledge-Based Systems Group\\
RWTH Aachen, Germany\\
\email{gerhard,delhibabu@kbsg.rwth-aachen.de}} \maketitle
%%%---------------------------------------------------------------------------------------------------------------------
\begin{abstract}
The dynamics of belief and knowledge is one of the major components
of any autonomous system  that should be able to incorporate new
pieces of information. In this paper, we argue that to apply
rationality result of belief dynamics theory to various practical
problems, it should be generalized in two respects: first of all, it
should allow a certain part of belief to be declared as immutable;
and second, the belief state need not be deductively closed. Such a
generalization of belief dynamics, referred to as base dynamics, is
presented, along with the concept of a generalized revision
algorithm for Horn knowledge bases. We show that Horn knowledge base
dynamics has interesting connection with kernel change and
abduction. Finally, we also show that both variants are rational in
the sense that they satisfy certain rationality postulates stemming
from philosophical works on belief dynamics.

\vspace{0.5cm}

\textbf{Keyword}: AGM, Belief Update, Horn Knowledge Base Dynamics,
Kernel Change, Abduction, View update.
\end{abstract}
%%---------------------------------------------------------------------------------------------------------------------
\section{Introduction}
Modeling intelligent agents' reasoning requires designing knowledge
bases for the purpose of performing symbolic reasoning. Among the
different types of knowledge representations in the domain of
artificial intelligence, logical representations stem from classical
logic. However, this is not suitable for representing or treating
items of information containing vagueness, incompleteness or
uncertainty, or Horn knowledge base evolution that leads the agent
to change his beliefs about the world.

When a new item of information is added to a Horn knowledge base,
inconsistency can result. Revision means modifying the Horn
knowledge base in order to maintain consistency, while keeping the
new information and removing (contraction) or not removing the least
possible previous information. In our case, update means revision
and contraction, that is insertion and deletion in database
perspective. Our previous work \cite{Arav1,Arav} makes connections
with contraction from knowledge base dynamics.

Our Horn knowledge base dynamics, is defined in two parts: an
immutable part (Horn formulae) and updatable part (literals) (for
definition and properties see works of Nebel \cite{Nebel} and
Segerberg \cite{Seg}). Knowledge bases have a set of integrity
constraints (see the definitions in later section). In the case of
finite knowledge bases, it is sometimes hard to see how the update
relations should be modified to accomplish certain Horn knowledge
base updates.

\begin{example} \label{E1} Consider a database with an (immutable) rule that a
staff member is a person who is currently working in the research
group under the chair. Additional (updatable) facts are  that
matthias and gerhard are group chairs, and delhibabu and aravindan
are staff members. We restricted that staff and chair names are
taken by her/his email id, and our integrity constraint is that each
research group has only one chair ie. $\forall x,y,z$ (y=x)
$\leftarrow$ group\_chair(x,y) $\wedge$ group\_chair(x,z).
\end{example}

\begin{center}
\underline {Immutable part}: staff\_chair(X,Y)$\leftarrow$
staff\_group(X,Z),group\_chair(Z,Y). \vspace{0.5cm}

\underline{Updatable part}: group\_chair(infor1,matthias)$\leftarrow$ \\
\hspace{2.4cm}group\_chair(infor2,gerhard)$\leftarrow$ \\
\hspace{2.6cm}staff\_group(delhibabu,infor1)$\leftarrow$ \\
\hspace{2.6cm}staff\_group(aravindan,infor2)$\leftarrow$ \\
\end{center}
Suppose we want to update this database with the information,
staff\_chair({delhiba\-bu},{aravindan}), that is

\begin{center}
staff\_chair(\underline{delhibabu},\underline{aravindan})$\leftarrow$
staff\_group(\underline{delhibabu},Z) $\bigwedge$
group\_chair(Z,\underline{aravindan})
\end{center}

If we are restricted to definite clauses, there is only one
plausible way to do this: delhibabu and aravindan belong to groups
infor1 and infor2, respectively, this updating means that we need to
delete (remove) matthias from the database and newly add (insert)
aravindan to the database (aravindan got promoted to the chair of
the research group infor1 and he was removed from research group
infor2). This results in an update that is too strong. If we allow
disjunctive information into the database, however, we can
accomplish the update by minimal adding wrt consistency
\begin{center}
staff\_group(\underline{delhibabu},infor1) $\lor$
group\_chair(infor1,\underline{aravindan})
\end{center}
and this option appears intuitively to be correct.

When adding new beliefs to the Horn knowledge base, if the new
belief is violating integrity constraints then belief revision needs
to be performed, otherwise, it is simply added. As we will see, in
these cases abduction can be used in order to compute all the
possibilities and it is \emph{not up to user or system} to choose
among them.

When dealing with the revision of a Horn knowledge base (both
insertions and deletions), there are other ways to change a Horn
knowledge base and it has to be performed automatically also.
Considering the information, change is precious and must be
preserved as much as possible. The \emph{principle of minimal
change} \cite{Herz,Schul} can provide a reasonable strategy. On the
other hand, practical implementations have to handle contradictory,
uncertain, or imprecise information, so several problems can arise:
how to define efficient change in the style of AGM \cite{Alch}; what
result has to be chosen \cite{Lak,Lobo,Nayak1}; and finally,
according to a practical point of view, what computational model to
support for Horn knowledge base revision has to be provided?

The rest of paper is organized as follows: First we start with
preliminaries in Section 2. In Section 3, we introduce knowledge
base dynamics along with the concept of generalized revision, and
revision operator for knowledge base. Section 4 studies the
relationship between knowledge base dynamics and abduction. In
Section 5, we discuss an important application of knowledge base
dynamics in providing an axiomatic characterization for insertion
view atoms to databases; and brief summary of the related works
nature of view update problem for incomplete to complete
information. In Section 6 we give brief overview of related works.
In Section 7 we make conclusions with a summary of our contribution
as well as a discussion of future directions of investigation. All
proofs can be found in the Appendix.
%%---------------------------------------------------------------------------------------------------------------------

\section{Preliminaries}

We consider a propositional language $\mathcal{L_P}$ defined from a
finite set of propositional variables $\mathcal{P}$ and the standard
connectives. We use lower case Roman letters $a, b, x, y,...$ to
range over elementary letters and the Greek letters $\varphi, \phi,
\psi, ...$ for propositional formulae. Sets of formulae are denoted
by upper case Roman letters $A,B, F,K, ....$. A literal is an atom
(positive literal), or a negation of an atom (negative literal).

For any formula $\varphi$, we write $E(\varphi)$ to mean the set of
the elementary letters that occur in $\varphi$. The same notation
also applies to a set of formulae. For any set $F$ of formulae,
$L(F)$ represents the sub-language generated by $E(F)$, i.e. the set
of all formulae $\varphi$ with $E(\varphi) \subseteq E(F)$.

Horn formulae are defined \cite{Delg} as follows:
\begin{enumerate}
\item[1.] Every $a \in \Phi$, $a$ and $\neg a$ are Horn clauses.
\item[2.] $a \leftarrow a_1 \land a_2 \land ... \land a_n$ is a Horn clause, where $n \geq 0$ and
$a, a_i \in \Phi$ ($1 \leq i \leq n$).
\item[3.] Every Horn clause is a Horn formula, $a$ is called head and $a_i$
is body of the Horn formula.
\item[4.] If $\varphi$ and $\psi$ are Horn formulae, so is $\varphi\land \psi$.
\end{enumerate}

A definite Horn clause is a finite set of literals (atoms) that
contains exactly one positive literal which is called the head of
the clause. The set of negative literals of this definite Horn
clause is called the body of the clause. A Horn clause is
non-recursive, if the head literal does not occur in its body. We
usually denote a Horn clause as head$\leftarrow$body. Let
$\mathcal{L_H}$ be the set of all Horn formulae with respect to
$\mathcal{L_P}$.

Formally, a finite Horn knowledge base $KB$ is defined as a finite
set of formula from language $\mathcal{L_{H}}$, and divided into
three parts: an immutable theory $KB_{I}$ is an Horn formulae
(head$\leftarrow$body), which is the fixed part of the knowledge;
updatable theory $KB_{U}$ is Horn clause (head$\leftarrow$); and an
integrity constraints $KB_{IC}$ is Horn clause ($\leftarrow$body).

\begin{definition} [Knowledge Base] \label{D1} Let KB be a finite set of Horn formulae from
language $\mathcal{L_{H}}$ called a Horn knowledge base with,
$KB=KB_{I}\cup KB_{U}\cup KB_{IC}$, $KB=KB_{I}\cap
KB_{U}=\varnothing$ and $KB=KB_{U}\cap KB_{IC}=\varnothing$.
\end{definition}

Working with deductively closed, infinite belief sets is not very
attractive from a computational point of view. The AGM approach to
belief dynamics is very attractive in its capturing the rationality
of change, but it is not always easy to implement either Horn
formula based partial meet revision, or model-theoretical revision.
In real application from artificial intelligence and database, what
is required is to represent the knowledge using a finite Horn
knowledge base. Further, a certain part of the knowledge is treated
as immutable and should not be changed.

Knowledge base change deals with situations in which an agent has to
modify its beliefs about the world, usually due to new or previously
unknown incoming information, also represented as formulae of the
language. Common operations of interest in Horn knowledge base
change are the expansion of an agent's current Horn knowledge base
KB by a given Horn clause $\varphi$ (usually denoted as
KB+$\varphi$), where the basic idea is to add regardless of the
consequences, and the revision of its current beliefs by $\varphi$
(denoted as KB * $\varphi$), where the intuition is to incorporate
$\varphi$ into the current beliefs in some way while ensuring
consistency of the resulting theory at the same time. Perhaps the
most basic operation in Horn knowledge base change, like belief
change, is that of contraction (AGM \cite{Alch}), which is intended
to represent situations in which an agent has to give up $\varphi$
from its current stock of beliefs (denoted as KB-$\varphi$).

\begin{definition} [Levi Identity] \label{D2} Let - be an AGM contraction
operator for KB. A way to define a revision is by using Generalized
Levi Identity:
\begin{center}
$KB*\alpha~=~(KB-\neg\alpha)\cup\alpha$
\end{center}
\end{definition}

Then, the revision can be trivially achieved by expansion, and the
axiomatic characterization could be straightforwardly obtained from
the corresponding characterizations of the traditional models
\cite{Fal}. The aim of our work is not to define revision from
contraction, but rather to construct and axiomatically characterize
revision operators in a direct way.

%%---------------------------------------------------------------------------------------------------------------------

\section{Knowledge base dynamics}

AGM \cite{Alch} proposed a formal framework in which
revision(contraction) is interpreted as belief change. Focusing on
the logical structure of beliefs, they formulate eight postulates
which a revision knowledge base (contraction knowledge base was
discussed in \cite{Arav}) has to verify.

\begin{definition} \label{D9} Let KB be a Horn knowledge base with an immutable part
$KB_{I}$. Let $\alpha$ and $\beta$ be any two Horn clauses from
$\mathcal{L_H}$. Then, $\alpha$ and $\beta$ are said to be
\emph{KB-equivalent} iff the following condition is satisfied:
$\forall$ set of Horn clauses E $\subseteq \mathcal{L_H}$:
$KB_{I}\cup E\vdash\alpha$ iff $KB_{I}\cup E\vdash\beta$.
\end{definition}

These postulates stem from three main principles: the new item of
information has to appear in the revised Horn knowledge base, the
revised base has to be consistent and revision operation has to
change the least possible beliefs. Now we consider the revision of a
Horn clause $\alpha$ wrt KB, written as $KB*\alpha$. The rationality
postulates for revising $\alpha$ from KB can be formulated.

\begin{definition} [Rationality postulates for Horn knowledge base
revision] \label{10}
\begin{enumerate}
\item[]\hspace{-0.6cm}(KB*1)\hspace{0.2cm}  \emph{Closure:} $KB*\alpha$ is a Horn knowledge base.
\item[]\hspace{-0.6cm}(KB*2)\hspace{0.2cm}  \emph{Weak Success:} if $\alpha$ is consistent with $KB_{I}\cup KB_{IC}$ then
$\alpha \subseteq KB*\alpha$.
\item[]\hspace{-0.6cm}(KB*3.1)  \emph{Inclusion:} $KB*\alpha\subseteq
Cn(KB\cup\alpha)$.
\item[]\hspace{-0.6cm}(KB*3.2)  \emph{Immutable-inclusion:} $KB_{I}\subseteq
Cn(KB*\alpha)$.
\item[]\hspace{-0.6cm}(KB*4.1)  \emph{Vacuity 1:} if $\alpha$ is
inconsistent with $KB_{I}\cup KB_{IC}$ then $KB*\alpha=KB$.
\item[]\hspace{-0.6cm}(KB*4.2)  \emph{Vacuity 2:} if $KB\cup \alpha \nvdash \perp$ then $KB*\alpha$ = $KB \cup
\alpha$.
\item[]\hspace{-0.6cm}(KB*5)\hspace{0.3cm}   \emph{Consistency:} if $\alpha$ is consistent with $KB_{I}\cup KB_{IC}$
 then $KB*\alpha$ consistent with $KB_{I}\cup KB_{IC}$.
\item[]\hspace{-0.6cm}(KB*6)  \hspace{0.2cm} \emph{Preservation:} If $\alpha$ and $\beta$ are
KB-equivalent, then $KB*\alpha \leftrightarrow KB*\beta$.
\item[]\hspace{-0.6cm}(KB*7.1)  \emph{Strong relevance:} $KB*\alpha\vdash \alpha$ If $KB_{I}\nvdash\neg\alpha$
\item[]\hspace{-0.6cm}(KB*7.2)  \emph{Relevance:} If $\beta\in KB\backslash KB*\alpha$,
then there is a set $KB'$ such that\\ $KB*\alpha\subseteq
KB'\subseteq KB\cup\alpha$, $KB'$ is consistent $KB_{I}\cup KB_{IC}$
with $\alpha$, but $KB' \cup \{\beta\}$ is inconsistent $KB_{I}\cup
KB_{IC}$ with $\alpha$.
\item[]\hspace{-0.6cm}(KB*7.3)  \emph{Weak relevance:} If $\beta\in KB\backslash KB*\alpha$,
then there is a set $KB'$ such that $KB'\subseteq KB\cup\alpha$,
$KB'$ is consistent $KB_{I}\cup KB_{IC}$ with $\alpha$, but $KB'
\cup \{\beta\}$ is inconsistent $KB_{I}\cup KB_{IC}$ with $\alpha$.
\end{enumerate}
\end{definition}

To revise $\alpha$ from KB, only those informations that are
relevant to $\alpha$ in some sense can be added (as example in the
introduction illustrates). $(KB*7.1)$  is very strong axiom allowing
only minimum changes, and certain rational revision can not be
carried out. So, relaxing this condition (example with more details
can be found in \cite{Arav}), this can be weakened to relevance.
$(KB*7.2)$ is relevance policy that still can not permit rational
revisions, so we need to go next step. With $(KB*7.3)$ the relevance
axiom is further weakened and it is referred to as
"core-retainment".

\subsection{Principle of minimal change}

Let a Horn knowledge base KB be a set of Horn formulae and $\psi$ is
a Horn clause such that $KB=\{\phi~|~\psi\vdash\phi\}$ is derived by
$\phi$. Now we consider the revision of a Horn clause $\alpha$ wrt
KB, that is $KB*\alpha$.

The principle of minimal change (PMC) leads to the definition of
orders between interpretations. Let $\mathcal{I}$ be the set of all
the interpretations and $Mod(\psi)$ be the set of models of $\psi$.
A pre-order on $\mathcal{I}$, denoted $\leq_{\psi}$ is linked with
$\psi$. The relation $<_{\psi}$ is defined from $\leq_{\psi}$ as
usual:
$$I<_{\psi}I'~\text{iff}~I\leq_{\psi}I'~\text{and}~I'\nleq_{\psi}I.$$

The pre-order $\leq_{\psi}$ is \it faithful \rm to $\psi$ if it
verifies the following conditions:

\begin{enumerate}
\item[1)] If $I,I'\in Mod(\psi)$ then $I<_{\psi}I'$ does not hold;
\item[2)] If $I\in Mod(\psi)$ and $I'\notin Mod(\psi)$ then
$I<_{\psi}I'$ holds;
\item[3)] if $\psi\equiv \phi$ then $\leq_{\psi}=\leq_{\phi}$.
\end{enumerate}

A minimal interpretation may thus be defined by:

$\mathcal{M}\subseteq \mathcal{I}$, the set of minimal
interpretations in $\mathcal{M}$ according to $\leq_{\psi}$ is
denoted $Min(\mathcal{M},\leq_{\psi})$. And $I$ is minimal in
$\mathcal{M}$ according to $\leq_{\psi}$, if $I\in\mathcal{M}$ and
there is no $I'\in \mathcal{M}$ such that $I'<_{\psi}I$.

Revision operation * satisfies the postulates (KB*1) to (KB*6) and
(KB*7.3) if and only if there exists a total pre-order $\leq_{\psi}$
such that: \begin{equation*} Mod(\psi *
\phi)=Min(Mod(\phi),\leq_{\psi}).
\end{equation*}

%%%----------------------------------------------------------------------------------

\section{Knowledge base dynamics and abduction}
We study the relationship between Horn knowledge base dynamics
(discussed in the previous section) and abduction, a well-known from
reasoning. This study helps to bring these two fields together, so
that abductive logic grammar procedure could be used to implement
revision. For this purpose, we use the concepts of generalized
kernel change (revision and contraction), an extension of kernel
contraction and revision introduced for belief bases. We first
observe that generalized kernel change coincides with that of Horn
knowledge base change (revision and contraction), and then we
process to show its relationship with abduction.

\subsection{Kernel revision system}

To revise a Horn formula $\alpha$ from a Horn knowledge base KB, the
idea of kernel revision is to \emph{keep at least} one element from
every inclusion-minimal subset of KB that derives $\alpha$. Because
of the immutable-inclusion postulate, no Horn formula from $KB_{I}$
can be deleted.

\begin{definition} [Kernel sets] \label{D14}
Let a Horn knowledge base KB be a set of Horn formulae, where
$\alpha$ is Horn clause. The $\alpha$-inconsistent kernel of KB,
noted by $KB\bot_{\bot}\alpha$, is the set of $KB'$ such that:

\begin{enumerate}
  \item $KB'\subseteq KB$ ensuring that $KB_{I}\subseteq KB'$ and $KB_{IC}\subseteq
  KB'$.
  \item $KB'\cup\alpha$ is inconsistent with $KB_{I}\cup KB_{IC}$ .
  \item For any KB'' such that $KB'' \subset KB'\subseteq KB$ then
  $KB''\cup\alpha$ is consistent with $KB_{I}\cup KB_{IC}$.
\end{enumerate}
\end{definition}

That is, given a consistent $\alpha$, $KB\bot_{\bot}\alpha$ is the
set of minimal KB-subsets inconsistent with $\alpha$.

\begin{example}\label{E4} Suppose that KB=\{$KB_{I}: p\leftarrow a\wedge b,p\leftarrow a,q\leftarrow
a\wedge b ;~KB_{U}: a\leftarrow, b\leftarrow; ~KB_{IC}: {\o}$\} and
$\alpha$= $\leftarrow p$. Then we have that:

$KB\bot_{\bot}\alpha$= \{$\{p\leftarrow a \wedge b\}, \{p\leftarrow
a \} $\}.
\end{example}

Revision by a Horn clause is based on the concept of a
$\alpha$-inconsistent-kernels. In order to complete the
construction, we must define a incision function that cuts in each
inconsistent-kernel.

\begin{definition}[Incision function] \label{D15} Let $KB$ be a set of Horn formulae.
$\sigma$ is a incision function for $KB$ if and only if, for all
consistent Horn clauses $\alpha$
\begin{enumerate}
  \item $\sigma(KB\bot_\bot \alpha) \subseteq \bigcup KB\bot_\bot
  \alpha$
  \item If $KB' \in KB\bot_\bot\alpha$ then
  $KB'\cap(\sigma(KB\bot_\bot\alpha))\neq 0$
\end{enumerate}
\end{definition}

\begin{definition} [Hitting set] \label{D16} A \emph{hitting set} H for $KB\bot_{\bot}\alpha$ is defined as
a set s.t. (i) $H\subseteq\bigcup(KB\bot_{\bot}\alpha)$, (ii) $H\cap
KB_{I}$ is empty and (iii) $\forall X \in KB\bot_{\bot}\alpha$,
$X\neq\emptyset$ and $X\cap KB_{U}$ is not empty, then $X\cap H
\neq\emptyset$.
\end{definition}

A hitting set is said to be \emph{maximal} when $H$ consists of all
updatable statements from $\bigcup(KB\bot_{\bot}\alpha)$ and
\emph{minimal} if no proper subset of $H$ is a hitting set for
$KB\bot_{\bot}\alpha$.

\begin{definition} [Generalized Kernel revision] \label{D17} An incision function for KB is a function s.t. for all
$\alpha$, $\sigma(KB\bot_{\bot}\alpha)$ is a hitting set for
$KB\bot_{\bot}\alpha$. An operator $*_{\sigma}$ for KB is a
generalized kernel revision defined as follows:

$$KB*_\sigma \alpha =\left\{ \begin{array}{cc} (KB\backslash
\sigma(KB\bot_{\bot}\alpha)\cup \alpha~~~~~&
\text{if}~\alpha ~\text{is consistent}~KB_{I}\cup KB_{IC}\\
KB&\text{otherwise.}
\end{array}\right.$$
\end{definition}

An operator $*_{\sigma}$ for KB is a generalized kernel revision iff
there is an incision function $\sigma$ for KB such that $KB*\alpha$
= $KB*_{\sigma}\alpha$ for all beliefs $\alpha$.

From the definition of hitting set, it is clear that when
$KB\vdash\neg\alpha$, $\alpha$ is the hitting set of
$KB\bot_\bot\alpha$. On the other hand, when $KB_{I}\vdash\alpha$,
the definition ensures that only updatable elements are inserted,
and $\alpha$ does follow from the revision. Thus, week success
(KB*2), immutable-inclusion(KB*3.2) and vacuity (KB*4.1) are
satisfied by generalized kernel revision of  $\alpha$ from KB.

\begin{example} \label{E5}
Given KB=\{$KB_{I}: p\leftarrow a\wedge b,p\leftarrow a,q\leftarrow
a\wedge b ;~KB_{U}: a\leftarrow, b\leftarrow; ~KB_{IC}: {\o}$ \},
$\alpha$= $\leftarrow p$ and $KB\bot_{\bot}\alpha = \{\{p\leftarrow
a\wedge b\}, \{ p\leftarrow a \}\}.$ We have two possible results
for the incision function  and its associated kernel revision
operator:
\begin{eqnarray*}
\sigma_1(KB\bot_{\bot}\alpha)&=&\{p\leftarrow a\wedge
b\}~\text{and}~
KB*_{\sigma_1} \alpha=\{\{\leftarrow a \}, \{ \leftarrow b \} \},\\
\sigma_2(KB\bot_{\bot}\alpha)&=&\{p\leftarrow a \}~\text{and}~
KB*_{\sigma_2} \alpha=\{\{\leftarrow a \} \}.
\end{eqnarray*}
Incision function $\sigma_2$ produces minimal hitting set for
$KB\bot_{\bot}\alpha$.

\end{example}

\begin{theorem} \label{T7} For every Horn knowledge base $KB$, $*_{\sigma}$ is a
generalized kernel revision function iff it satisfies the postulates
(KB*1) to (KB*6) and (KB*7.3).
\end{theorem}

\subsection{Relationship with abduction}
The relationship between Horn knowledge base dynamics and abduction
was introduced by the philosopher Pierce (see \cite{Alis}). We show
how abduction grammar could be used to realize revision with
immutability condition. A special subset of literal (atoms) of
language $\mathcal{L_{H}}$, \emph{abducibles} Ab, are designated for
abductive reasoning. An abductive framework $\langle P,Ab\rangle$
stands for a theory P, which is a set of Horn formulae from
$\mathcal{L_{H}}$, with possible hypotheses $Ab$. An abductive
framework for a knowledge base $KB=KB_{I}\cup KB_{U} \cup KB_{IC}$
can be given as follows:
\begin{eqnarray*}
P=KB_{I}\cup \{&&\hspace{-0.2cm}\alpha\leftrightarrow\beta |
\alpha~\text{\rm is a Horn clause in}~ KB_{U} ~\text{\rm and}~
\beta~\text{\rm is an abducible }\\
&& \text{\rm  from Ab that does not appear in}~ KB \}.
\end{eqnarray*}

\begin{definition}[Minimal abductive explanation] \label{D20}
Let KB be a Horn knowledge base and $\alpha$ an observation to be
explained. Then, for a set of abducibles $(KB_{U})$, $\Delta$ is
said to be an abductive explanation wrt $KB_{I}$ iff $KB_{I}\cup
\Delta\vdash \alpha$. $\Delta$ is said to be \emph{minimal} wrt
$KB_{I}\cup KB_{IC}$ iff no proper subset of $\Delta$ is an
abductive explanation for $\alpha$, i.e. $\nexists\Delta^{'}$ s.t.
$KB_{I}\cup\Delta^{'}\vdash\alpha$.
\end{definition}

Since an incision function is adding and removing only updatable
elements from each member of the kernel set, to compute a
generalized revision of $\alpha$ from KB, we need to compute only
the abduction in every $\alpha$-kernel of KB. So, it is now
necessary to characterize precisely the abducibles present in every
$\alpha$-kernel of KB. The notion of minimal abductive explanation
is not enough to capture this, and we introduce locally minimal and
KB-closed abductive explanations.

\begin{definition}[Local minimal abductive explanations] \label{D21}
Let $(KB_I\cup KB_U')$ be a smallest subset of $KB_{U}$, s.t
$\Delta$ an minimal abductive explanation of $\alpha$ wrt $(KB_I\cup
KB_U')$ (for some $\Delta$). Then $\Delta$ is called local minimal
for $\alpha$ wrt $KB_{U}$.
\end{definition}

\begin{note} \label{N2}
Let $(KB_I\cup KB_U)\in(\{\Delta^{+},\Delta^{-}\})$. Here
$\Delta^{+}$ refers to admission Horn knowledge base (positive
atoms) and $\Delta^{-}$ refers to denial Horn knowledge
base(negative atoms) wrt given $\alpha$. Then problem of abduction
is to explain $\Delta$ with abducibles $(KB_{U})$, s.t. $(KB_I\cup
KB_U)\cup\Delta^{+}\cup\Delta^{-}\vdash \alpha$ and $(KB_I\cup
KB_U)\cup\Delta^{+}\models\alpha\cup\Delta^{-}$ are both consistent
with IC.
\end{note}

%--------------------------------------------------------------------------------------------------------------

\subsection{Generalized revision algorithm}

The problem of Horn knowledge base revision is concerned with
determining how a request to change can be appropriately translated
into one or more atoms or literals. We give new generalized revision
algorithm. It is enough to compute all the KB-locally minimal
abduction explanations for $\alpha$ wrt $KB_I \cup KB_U \cup
KB_{IC}$. If $\alpha$ is consistent with KB then well-known
abductive procedure to compute an abductive explanation for $\alpha$
wrt $KB_{I}$ could be used to compute kernel revision\\

\textbf{Reasoning about Abduction and Deduction}

\begin{definition}[\text{[51]}] \label{D23}
Let KB=($KB_I,KB_U,KB_{IC}$) be a knowledge base, $T$ is updatable
part from KB. We define abduction framework $\langle
KB^{BG},KB^{Ab},IC\rangle$. After Algorithm 1 is executed, $u$ is
derived part from $KB'$. The abduction explanation for $u$ in
$\langle KB_I\cup KB_U^*, KB_{IC}\rangle$ is any set $T_i$, where
$T_i \subseteq KB^{Ab}$ such that: $KB_I\cup KB_U^* \cup T \models
u$.

An explanation $T_i$ is minimal if no proper subset of $T_i$ is also
an explanation, i.e. if it does not exist any explanation $T_j$  for
$u$ such that $T_j \subset T_i$
\end{definition}

\begin{definition}[\text{[51]}] \label{D24}
Let KB=($KB_I,KB_U,KB_{IC}$) be a knowledge base, $T$ is updatable
part from KB. After Algorithm 1 is executed, $u$ is derived part
from $KB'$. The deduction consequence on $u$ due to the application
of $T$, $KB_I\cup KB_U^* \cup T\cup u$ is the answer to any
question.
\end{definition}

$$\begin{array}{cc}\hline \text{\bf Algorithm 1} & \hspace{-4cm}
\text{\rm Generalized revision algorithm}\\\hline \text{\rm Input}:&
\hspace{-0.6cm}\text{\rm A Horn knowledge base}~ KB=KB_{I}\cup KB_{U}\cup KB_{IC}\\
&\text{\rm and a Horn clause}~
\alpha~ \text{\rm to be revised.}\\
\text{\rm Output:} & \text{\rm A new Horn knowledge base
}~KB'=KB_{I}\cup
KB_{U}^*\cup KB_{IC},\\
&\text{s.t.}~ KB'\text{\rm is
a generalized revision}~ \alpha~\text{\rm to KB.}\\
\text{\rm Procedure}~KB(KB,\alpha)&\\
\text{\rm begin}&\\
~~1.&\hspace{-0.5cm}\text{\rm Let V:=}~\{c\in KB_{IC}~|~ KB_I\cup
KB_{IC}~\text{\rm inconsistent
with}~\alpha~\text{\rm wrt}~c\}\\
&P:=N:=0~\text{\rm and}~KB'=KB\\
~~2.&\text{\rm While}~(V\neq 0)\\
&\text{\rm select a subset}~V'\subseteq V\\
&\text{\rm For each}~v\in~V',~\text{\rm select a literal to be}\\
&\hspace{-0.1cm}\text{\rm remove (add to N) or a literal to be added(add to P)}\\
&\text{\rm Let KB}~:=KR(KB,P,N)\\
&\hspace{-0.3cm}\text{\rm Let V:=}~\{c\in KB_{IC}~|~ KB_I~\text{\rm
inconsistent
with}~\alpha~\text{\rm wrt}~c\}\\
&\hspace{-0.7cm}\text{\rm return}\\
~~3.&\text{\rm Produce a new Horn knowledge base}~KB'\\
\text{\rm end.}&\\ \hline
\end{array}$$

\vspace{0.5cm}

$$\begin{array}{cc}\hline
\text{\bf Algorithm 2} \label{A2} &\\
\text{\rm Procedure}~
KR(KB,\Delta^{+},\Delta^{-})&\\
\text{\rm begin}&\\
1.&\hspace{-1.6cm}\text{\rm Let}~ P :=\{ e \in \Delta^{+} |~
KB_I\not\models e\} ~\text{\rm and}~ N :=\{ e \in \Delta^{-}
 |~KB_I\models e\}\\
2.&\text{\rm While}~(P\neq 0)~\text{\rm or}~(N\neq 0)\\
&\text{\rm select a subset}~P'\subseteq P~ or ~N'\subseteq N \\
&\hspace{-1.7cm}\text{\rm Construct a set}~S_1=\{X~|~X~\text{\rm is
a KB-closed
locally}\\
&\text{\rm minimal abductive
wrt P explanation for}~\alpha~\text{\rm wrt}~KB_{I}\}.\\
&\hspace{-1.7cm}\text{\rm Construct a set}~S_2=\{X~|~X~\text{\rm is
a KB-closed
locally}\\
&\text{\rm  minimal abductive wrt N explanation for}~\alpha~\text{\rm wrt}~KB_{I}\}.\\
3.&\text{\rm Determine a hitting set}~\sigma (S_1) \text{\rm ~and}~\sigma (S_2)\\
&\hspace{-5.5cm}
\text{\rm If}~((N=0)~and~(P\neq0))\\
&\hspace{-1cm}\text{\rm Produce}~KB'=KB_{I}\cup \{(KB_{U} \cup \sigma (S_1)\}\\
&\hspace{-8.8cm}
\text{\rm else}\\
&\text{\rm Produce}~KB'=KB_{I}\cup \{(KB_{U}\backslash
\sigma(S_2) \cup \sigma (S_1)\}\\
&\hspace{-8.5cm}
\text{\rm end if}\\
&\hspace{-5.5cm}
\text{\rm If}~((N\neq0)~\text{\rm and}~(P=0))\\
&\hspace{-1.2cm}\text{\rm Produce}~KB'=KB_{I}\cup
\{(KB_{U}\backslash
\sigma(S_2)\}\\
&\hspace{-8.8cm}
\text{\rm else}\\
&\text{\rm Produce}~KB'=KB_{I}\cup \{(KB_{U}\backslash
\sigma(S_2) \cup \sigma (S_1)\}\\
&\hspace{-8.5cm}
\text{\rm end if}\\
4.& \hspace{-1.5cm}
\text{\rm return}~ KB'\\
\text{\rm end.}&\\ \hline
\end{array}$$

\begin{theorem} \label{T8} Let KB be a Horn knowledge base and $\alpha$
is Horn formula.
\begin{enumerate}
  \item If Algorithm 1 produced KB'as a result of revising $\alpha$
  from KB, then KB' satisfies all the rationality postulates (KB*1) to
(KB*6) and (KB*7.3).
  \item Suppose $KB''$ satisfies all these rationality postulates
  for revising $\alpha$ from KB, then $KB''$ can be produced by Algorithm 1.
\end{enumerate}
\end{theorem}

%%%--------------------------------------------------------------------------------

\section{Application: View updates in database}
An important application of knowledge base dynamics, discussed in
the previous section, is in providing an axiomatic characterization
of view updates in deductive and relational databases. A \it
definite deductive database \rm DDB consists of two parts: an \it
intensional database \rm IDB ($KB_{I}$), a set of definite program
clauses; and an \it extensional database \rm EDB ($KB_{U}$), a set
of ground facts. The intuitive meaning of DDB is provided by the \it
Least Herbrand model semantics \rm and all the inferences are
carried out through \it SLD-derivation. All the predicates that are
defined in IDB are referred to as \it view predicates\rm and those
defined in EDB are referred to as \it base predicates. \rm Extending
this notion, an atom(literals) with a view predicate is said to be a
\it view atom(literals), \rm and similarly an atom(literals) with
base predicate is a \it base atom(literals). \rm Further, we assume
that IDB does not contain any unit clauses and that predicates
defined in a given DDB are both view and base predicates.
%
%\begin{figure}
%\begin{center}
%   \includegraphics[height=2cm,angle=0]{4}
%   \caption{View Atom}
%   \end{center}
%\end{figure}

Two kinds of view updates can be carried out on a DDB: An
atom(literals), that does not currently follow from DDB, can be \it
inserted; \rm or an atom(literals), that currently follows from DDB,
can be \it deleted \cite{Arav1,Arav}. \rm In this paper, we consider
only insertion an atom(literals) from a DDB. When an atom(literals)
$A$ is to be inserted, the view update problem is to delete only
some relevant EDB facts and then to insert, so that the modified EDB
together with IDB will satisfy the insertion of $A$ from DDB. As
motivated in the introduction, our concern now is to discuss the
rationality of view update, and provide an axiomatic
characterization for it. This axiomatic characterization can be seen
as a declarative semantics for view updates in deductive databases.

Note that DDB can be considered \cite{Min,Sie} as a knowledge base
to be revised. The IDB is the immutable part of the knowledge
database, while the EDB forms the updatable part. Every base literal
is an abducible, but since we deal only with definite databases, we
require only positive abducibles. In general, it is assumed that a
language underlying a DDB is fixed and the semantics of DDB is the
least Herbrand model over this fixed language. Therefore, the DDB is
practically a shorthand of its ground instantiation\footnotemark
\footnotetext{a ground instantiation of a definite program $P$ is
the set of clauses obtained by substituting terms in the Herbrand
Universe for variables in $P$ in all possible ways}, written as
$IDB_G$. Thus, a DDB represent a knowledge base where the immutable
part is given by $IDB_{G}$ and updatable part is EDB. Hence, the
rationality postulates (KB*1) to (KB*6) and (KB*7.3) provide an
axiomatic characterization for inserting a view atom(literals) $A$
to a definite database DDB, and a generalized insertion of $A$ to
DDB achieves deletion of $A$ from DDB.

As observed by Kowalski \cite{Kow}, logic can provide a conceptual
level of understanding of relational databases, and hence
rationality postulates (KB*1) to (KB*6) and (KB*7.3)can provide an
axiomatic characterization for view insertion in relational
databases too. A relational database together with its view
definitions can be represented by a definite deductive database (EDB
representing tuples in the database and IDB representing the view
definitions), and so same algorithm can be used to insert view
extensions from relational and deductive databases.

But before discussing the rationality postulates and algorithm, we
want to make it precise, how a relational database, along with
operations on relations, can be represented by definite deductive
database. We assume the reader is familiar with relational database
concepts. \it A relation scheme \rm $R$ can be thought of as a base
predicate whose arguments define the \it attributes \bf A \rm of the
scheme. Its \it relational extension \rm $r$, is a finite set of
base atoms $R(\mathbb A)$ containing the predicate $R$. A \it
database schema \rm consists of finite collection of relational
schemes $<R_1,\ldots,R_n>$, and a \it relational database \rm is a
specific extension of database schema, denoted as
$<r_1,\ldots,r_n>$. In our context, relational database can be
represented by $EDB=\bigcup_{i=1,\ldots,n}R_i(\mathbb{A}_i)$.

\it Join \rm is a binary operator for combining two relations. Let
$r$ and $s$ be two relational extensions of schema $R$ (with
attributes $\mathbb R$) and $S$ (with attributes $\mathbb S$),
respectively. Let $\mathbb T=\mathbb R\cup\mathbb S$. The join of
$r$ and $s$, written as $r\otimes s$, is the relational extension
$q(\mathbb T)$ of all tuples $t$ over $\mathbb T$ such that there
are $t_r\in r$ and $t_s\in s$, with $t_r=t(\mathbb R)$ and
$t_s=t(\mathbb S)$. Join can be captured by a constraint clause
$Q(\mathbb T)\leftarrow R(\mathbb R), S(\mathbb S)$. Our integrity
constraint (IC) is that each research group has only one chair i.e.
$\forall x,y,z$ (y=x) $\leftarrow$ group\_chair(x,y) $\wedge$
group\_chair(x,z) (see definition and properties of similarity in
works of Christiansen \cite{Chris1} and Godfrey \cite{God}).

\begin{example} \label{E11} Let us consider two relational schemes $R$ and $S$
from Example 1, with attributes $R=\{Group,Chair\}$ and $S=\{Staff,
Group\}$.Consider the following extensions $r$ and $s$:
\begin{center}
$\begin{array}{c|cc} \text{\rm s}&\text{Staff}&\text{Group}\\\hline
&\text{\rm delhibabu}&\text{\rm infor1}\\
&\text{\rm aravindan}&\text{\rm infor2}\\
\end{array}$
$~~~~~~~\begin{array}{c|cc} \text{\rm
r}&\text{Group}&\text{Chair}\\\hline
&\text{\rm infor1}&\text{\rm matthias}\\
&\text{\rm infor2}&\text{\rm gerhard}\\
\end{array}$
\end{center}

\begin{center} \text{\bf Tab. 1. \rm Base table for $s$ and $r$}\end{center}
\end{example}

The following rule, $T(Staff,Group,Chair)\leftarrow$
$S(Staff,Group),R(Group,Chair)$\\

represents the join of $s$ and $r$, which is given as:
$$\begin{array}{c|ccc}
s\otimes r&Staff&Group&Chair\\\hline
&\text{\rm delhibabu}&\text{\rm infor1}&\text{\rm matthias}\\
&\text{\rm aravindan}&\text{\rm infor2}&\text{\rm gerhard}\\
\end{array}$$
\begin{center} \text{\bf Tab. 2. \rm $s\otimes r$ }\end{center}

To sum up, we showed how relational database and operators on
relations can be conceptually captured by definite deductive
databases. All solutions translate \cite{Mota} a view update request
into a \textbf{transaction combining insertions and deletions of
base relations} for satisfying the request. Further, a definite
deductive database can be considered as a knowledge base, and thus
rationality postulates and insertion algorithm of the previous
section can be applied for view updates in database.

\subsection{View insertion algorithm}

Since relational and definite deductive databases can be considered
as knowledge bases, and inserting a view atom(literals) (tuple) $A$
can be considered as revision of $A$, a specific instance of
Algorithm 1 can be used to compute insertion of a view
atom(literals) to a database. In fact, we have to discuss how to
compute all DDB-closed locally minimal abductive explanations for
$A$ wrt $IDB_G$. As expected, these abductive explanations can be
computed using deduction trees, and the process is discussed in the
sequel.

$$\begin{array}{cc}\hline
\text{\bf Algorithm 3} & \text{\rm Algorithm to compute all
DDB-closed locally minimal}\\  &\text{\rm abductive explanation of
an atom(literals)}\\\hline \text{\rm Input}:& \text{\rm A definite
deductive database}~DDB=IDB\cup EDB\cup IC~\text{\rm
an literals}\\
&\mathcal{A}\\
\text{\rm Output}:&\text{\rm Set of all DDB-closed locally minimal
abductive explanations}\\
&\text{\rm for}~\mathcal{A}~\text{\rm wrt}~IDB_{G}\\
\text{\rm begin}&\\
~~1.&\text{\rm Let}~ V :=\{ c\in IC~|~IDB\cup IC~\text{\rm
inconsistent
with}~\mathcal{A}~\text{\rm wrt}~c~\}\\
&\text{\rm While}~(V\neq 0)\\
&\hspace{-1.9cm}\text{\rm Construct a complete SLD-tree for} \leftarrow\mathcal{A}~\text{\rm wrt DDB.}\\
&\hspace{-0.8cm}\text{\rm For every successful branch $i$: construct}~\Delta_{i}=\{D~|~D \in EDB\\
&\text{\rm and D is used as an input clause in branch $i$}\}\\
%&\text{\rm Construct a complete SLD-tree for} \leftarrow\mathcal{A}~\text{\rm wrt DDB.}\\
&\hspace{-0.3cm}\text{\rm For every unsuccessful branch $j$: construct}~\Delta_{j}=\{D~|~D \in EDB\\
&\text{\rm and D is used as an input clause in branch $j$}\}\\
&\text{\rm Produce set of all}~\Delta_{i}~\text{\rm
and}~\Delta_{j}~\text{\rm computed in
 the previous step}\\
 &\text{\rm as the result.}\\
% &\text{\rm Let}~ V :=\{ c\in IC~|~IDB\cup IC~\text{\rm inconsistent
%with}~\mathcal{A}~\text{\rm wrt}~c~\}\\
&\hspace{-0.7cm}\text{\rm return}\\
~~2.&\text{\rm Produce all DDB-closed locally minimal abductive}\\
&\text{\rm explanations in}~\Delta_{i}~\text{\rm and}~\Delta_{j}\\
 \text{\rm end.}\\\hline
\end{array}$$

$$\begin{array}{cc}\hline
\text{\bf Algorithm 4} & \text{\rm Algorithm to compute all
DDB-closed locally minimal}\\  &\text{\rm abductive explanation of
an atom(literals)}\\\hline \text{\rm Input}:& \text{\rm A definite
deductive database}~DDB=IDB\cup EDB\cup IC~\text{\rm
an literals}\\
&\mathcal{A}\\
\text{\rm Output}:&\text{\rm Set of all DDB-closed locally minimal
abductive explanations}\\
&\text{\rm for}~\mathcal{A}~\text{\rm wrt}~IDB_{G}\\
\text{\rm begin}&\\
~~1.&\hspace{-1.9cm}\text{\rm Construct a complete SLD-tree for} \leftarrow\mathcal{A}~\text{\rm wrt DDB.}\\
&\hspace{-0.8cm}\text{\rm For every successful branch $i$: construct}~\Delta_{i}=\{D~|~D \in EDB\\
&\text{\rm and D is used as an input clause in branch $i$}\}\\
%&\text{\rm Construct a complete SLD-tree for} \leftarrow\mathcal{A}~\text{\rm wrt DDB.}\\
&\hspace{-0.3cm}\text{\rm For every unsuccessful branch $j$: construct}~\Delta_{j}=\{D~|~D \in EDB\\
&\text{\rm and D is used as an input clause in branch $j$}\}\\
~~2.&\text{\rm Let}~ V :=\{ c\in IC~|~IDB\cup IC~\text{\rm
inconsistent with}~\mathcal{A}~\text{\rm wrt}~c~\}\\
&\text{\rm While}~(V\neq 0)\\
&\text{\rm Produce set of all}~\Delta_{i}~\text{\rm
and}~\Delta_{j}~\text{\rm is consistent with IC}\\
 &\text{\rm as the result.}\\
&\hspace{-0.7cm}\text{\rm return}\\
&\text{\rm Produce all DDB-closed locally minimal abductive}\\
&\text{\rm explanations in}~\Delta_{i}~\text{\rm and}~\Delta_{j}\\
 \text{\rm end.}\\\hline
\end{array}$$

An update request U = B, where B is a set of base facts, is not true
in KB. Then, we need to find a transaction $T=T_{ins} \cup T_{del}$,
where $T_{ins} (\Delta_i)$ (resp. $T_{del}(\Delta_j)$) is the set of
facts, such that U is true in $DDB'=((EDB - T_{del} \cup T_{ins})
\cup IDB \cup IC)$. Since we consider definite deductive databases,
SLD-tree can be used to compute the required abductive explanations.
The idea is to get all EDB facts used in a SLD-derivation of $A$ wrt
DDB, and construct that as an abductive explanation for $A$ wrt
$IDB_G$.

There are two ways to find minimal elements (insertion and deletion)
with integrity constraints. Algorithm 3 first checks consistency
with integrity constraints and then reduces steps with abductive
explanation for $A$ . Algorithm 4 is doing \emph{vice versa}, but
both algorithm outputs are similar.

 Unfortunately, this algorithm
does not work as intended for any deductive database, and a counter
example is produced below. Thus, general algorithms 3 and 4 produced
some unexpected sets in addition to locally minimal abductive
explanations.

\begin{example} \label{E12}
Consider a deductive database DDB as follows:

$$\begin{array}{cccccc} IDB:&p\leftarrow
a\wedge e&\hspace{0.5cm}EDB:&a\leftarrow&\hspace{1.2cm}IC:&\leftarrow b\\
&q\leftarrow a\wedge f&&e\leftarrow&&\\
&p\leftarrow b\wedge f&&f\leftarrow&&\\
&q\leftarrow b\wedge e&&&&\\
&\hspace{-0.6cm}p\leftarrow q&&&&\\
&\hspace{-0.6cm}q\leftarrow a&&&&\end{array}$$

We need to insert $p$. First, we check consistency with IC and after
we find $\Delta_{i}$ and $\Delta_{j}$ via tree deduction.

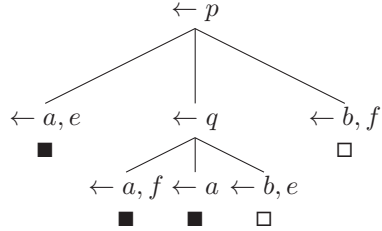
\Tree[ {$\leftarrow a,e$\\$\blacksquare$} [.$\leftarrow q$
{$\leftarrow a,f$\\$\blacksquare$} {$\leftarrow a$\\$\blacksquare$}
{$\leftarrow b,e$\\$\Box$} ].$\leftarrow q$ {$\leftarrow
b,f$\\$\Box$} ].$\leftarrow p$

\end{example}

From Algorithm 3 it is easy to conclude which branches are
consistent wrt IC (shown on tree by $\blacksquare$). For the next
step, we need to find minimal accommodate and denial literal with
wrt to $p$. The subgoals of the tree are $\leftarrow a,e$ and
$\leftarrow a,f$, which are minimal tree deductions of only facts.
Clearly, $\Delta_{i} =\{a,e,f\}$ and $\Delta_{j} =\{b\}$ with
respect to IC, are the only locally minimal abductive explanations
for $p$ wrt $IDB_G$, but they are not locally minimal explanations.

From Algorithm 4, the subgoals of the tree are $\leftarrow a,e$,
$\leftarrow a,f$, $\leftarrow b,f$ and $\leftarrow b,e$. Clearly,
$\Delta_{i} =\{a,b,e,f\}$ and $\Delta_{j} =\{a,e,f\}$. In the next
step, we check consistency with IC. $\Delta_{i}$ and $\Delta_{j}$
are only locally minimal abductive explanations for $p$ wrt $IDB_G$,
but they are not locally minimal explanations (more explanations can
be found in \cite{Lu}).

The program is clear due to the unwanted recursion $p\leftarrow
a\wedge b, p\leftarrow a$. Will the algorithm work as intended if we
restrict ourselves to acyclic program \cite{Arav} that excludes such
loop? One would expect a positive answer, but unfortunately still
some unwanted sets may be produced as the following example
highlights.

So, even for acyclic program, algorithms 3 and 4 do not work as
intended (that is to generate all and only the DDB-close locally
minimal abductive explanations). Does this mean that generalized
revision can not be carried out for database in general?. Probably
we should approach the problem from different perspective. We have
seen that algorithms 3 and 4 may compute some unwanted sets in
addition to the required ones. What exactly are those sets? Is it
possible to characterize them? The following lemma answers these
questions.

\begin{lemma} \label{l2} Let $DDB=IDB\cup EDB \cup IC$ be a definite deductive database
and $A$ an atom(literals). Let $S$ be the set of all DDB-closed
locally minimal abductive explanations for $A$ wrt $IDB_G$. Let $S'$
be the set of explanations returned by algorithms 3 and 4 given DDB
and $A$ as inputs. Then, the following propositions hold:
\begin{enumerate}
\item[1.] $S\subseteq S'$.
\item[2.] $\forall \Delta'(\Delta'\in\Delta_i \cup \Delta_j) \in S'$: $\exists \Delta\in S$ s.t. $\Delta\subset \Delta'$.
\item[3.] Suppose DDB is resticted to be acyclic then:
$\forall \Delta'\in S'$: $\Delta'\subset \bigcup S$.
\end{enumerate}
\end{lemma}

Having characterized what exactly is computed by algorithms 3 and 4,
we now proceed to show that algorithms 5 and 6 are useful for view
insertion. The key to the solution is the following lemma, which
established the preservable of hitting set computation among two
sets.

\begin{lemma} \label{l3}
\begin{enumerate}
\item[1.] Let $S$ be a set of sets, and $S'$ another set s.t.
$S\subseteq S'$ and every member of $S'\backslash S$ contains an
element of $S$. Then, a set $H$ is minimal hitting set for $S$ iff
it is a minimal hitting set for $S'$.
\item[2.] Let $S$ be a set of sets, and $S'$ another set s.t.
$S\subseteq S'$ and for every member $X$ of $S'\backslash S$: $X$
contains a member of $S$ and $X$ is contained in $\bigcup S$. Then,
a set $H$ is a hitting set for $S$ iff it is a hitting set for $S'$.
\end{enumerate}
\end{lemma}

Thus algorithms 3 and 4 in conjunction with an algorithm to compute
minimal hitting set can be used to compute partial meet revision
(defined in section 4.1) of $A$ from DDB.

$$\begin{array}{cc}\hline
\text{\bf Algorithm 5} &  \text{\rm Partial meet revision for
definite deductive database}\\\hline \text{\rm Input}:& \text{\rm A
definite deductive database}~ DDB=IDB\cup EDB\cup IC~~\text{\rm an
literals}~\mathcal{A}\\
\text{\rm Output:}&\text{\rm A Partial meet revision
of}~\mathcal{A}~\text{\rm from DDB.}\\
\text{\rm begin}&\\
~~1.&\text{\rm Let}~ V :=\{ c\in IC~|~IDB\cup IC~\text{\rm
inconsistent
with}~\mathcal{A}~\text{\rm wrt}~c~\}\\
&\text{\rm While}~(V\neq 0)\\
~~2.&\text{\rm Construct a complete SLD-tree for}~\leftarrow \mathcal{A}~\text{\rm wrt DDB.}\\
~~3.&\hspace{-0.7cm}\text{\rm For every successful branch $i$:construct}~\Delta_{i}=\{D~|~D\in EDB \}\\
&\text{\rm and D is used as an input clause in branch $i$}.\\
&\text{\rm  Let there be \it m \rm such sets.}\\
&\text{\rm Let}~ E^{*}=\{\{D_{1},\ldots,D_{m}\}|D_{i}\in \Delta_{i}\}\\
&\text{\rm Let E be a inclusion-minimal set among}~E*,\text{\rm
 i.e.}~ \nexists E'\in E^{*}\\
 &\text{\rm s.t.}~E'\subset E.\\
~~4.&\text{\rm For every unsuccessful branch $j$:construct}~\Delta_{j}=\{D~|~D\in EDB \}\\
&\text{\rm and D is used as an input clause in branch $j$}.\\
&\text{\rm Let there be \it m \rm such sets.}\\
&\text{\rm Let}~ F^{*}=\{\{D_{1},\ldots,D_{m}\}|D_{j}\in \Delta_{j}\}\\
&\text{\rm Let F be a inclusion-maximum set among}~F*,\text{\rm
 i.e.}~ \nexists F'\in F^{*}\\
 &\text{\rm s.t.}~F'\subseteq F.\\
 &\text{\rm Let}~ V :=\{ c\in IC~|~IDB\cup IC~\text{\rm inconsistent
with}~\mathcal{A}~\text{\rm wrt}~c~\}\\
&\hspace{-0.7cm}\text{\rm return}\\
~~5.&\text{\rm Produce}~DDB\backslash F \cup E ~\text{\rm as the
result.}\\
\text{\rm end.}&\\\hline
\end{array}$$

$$\begin{array}{cc}\hline
\text{\bf Algorithm 6} &  \text{\rm Generalized revision for acyclic
definite}\\
&\text{\rm deductive database}\\\hline \text{\rm Input}:& \text{\rm
An acyclic definite deductive database}~ DDB=IDB\cup EDB\cup IC \\
&\text{\rm
an literals}~\mathcal{A}\\
\text{\rm Output:} & \text{\rm A generalized revision
of}~\mathcal{A}~\text{\rm from DDB.}\\
\text{\rm begin}&\\
~~1.&\text{\rm Let}~ V :=\{ c\in IC~|~IDB\cup IC~\text{\rm
inconsistent
with}~\mathcal{A}~\text{\rm wrt}~c~\}\\
&\text{\rm While}~(V\neq 0)\\
~~2.&\text{\rm Construct a complete SLD-tree for}
\leftarrow \mathcal{A}~\text{\rm wrt DDB.}\\
~~3.&\hspace{-1.5cm}\text{\rm For every successful branch $i$:construct}~\Delta_{i}=\{D~|~D\in EDB \}\\
&\text{\rm and D is used as an input clause in branch $i$}.\\
&\hspace{-0.8cm}\text{\rm Construct a hitting set D for
all}~\Delta_{i}\text{\rm 's
computed in the previous step.}\\
~~4.&\hspace{-1.1cm}\text{\rm For every unsuccessful branch $j$:construct}~\Delta_{j}=\{D~|~D\in EDB \}\\
&\text{\rm and D is used as an input clause in branch $j$}.\\
&\hspace{-0.7cm}\text{\rm Construct a hitting set D for
all}~\Delta_{j}\text{\rm 's
computed in the previous step.}\\
 &\text{\rm Let}~ V :=\{ c\in
IC~|~IDB\cup IC~\text{\rm inconsistent
with}~\mathcal{A}~\text{\rm wrt}~c~\}\\
&\hspace{-0.7cm}\text{\rm return}\\
~~5.&\text{\rm Produce}~DDB\backslash F \cup E ~\text{\rm as the
result.}\\
\text{\rm end.}&\\\hline
\end{array}$$

When DDB is acyclic, generalized revision of $A$ from DDB can be
obtained by Algorithm 6. Observe that the first two steps of
Algorithm 5 are same as those of algorithms 3 and 4, and we have
already established what exactly are computed by them. Steps 3 and 4
clearly compute a minimal hitting set and as established by lemma
\ref{l2} and lemma \ref{l3}, this algorithm produces a partial meet
contraction of $A$ from DDB. This result is formalized below.

\begin{theorem} \label{T14} Let DDB be a definite deductive database and $A$ an
atom(literals) to be inserted. Then DDB' is a result of algorithm 5
given DDB and $A$ as inputs, iff DDB' is a partial meet revision of
$A$ from DDB, satisfying the postulates (KB*1) to (KB*6) and
(KB*7.1).
\end{theorem}

We proceed to present Algorithm 6 to compute generalized revision
for definite deductive database. As observed before, this is not
possible in general, but for a restricted case of acyclic program.

\begin{theorem} \label{T15} Let DDB be a definite deductive database and $A$ an
atom(literals) to be inserted. Then DDB' is a result of algorithm 6
given DDB and $A$ as inputs, iff DDB' is a generalized revision of
$A$ from DDB, satisfying the postulates (KB*1) to (KB*6) and
(KB*7.3).
\end{theorem}

Algorithms 5 and 6 are inefficient, as they need to build a complete
SLD-tree. Unfortunately, any rational algorithm for insertion can
not avoid  constructing complete SLD-trees. If these algorithms are
changed to extract input clauses from incomplete SLD-derivation,
then the new algorithm should check the derivability of an
atom(literals) from a deductive database, before any insertion is
carried out(otherwise, success can not be satisfied). Checking
derivability is also computationally expensive and more then that,
weak relevance policy (KB*7.3) will not be satisfied in general.
Finally, any rational algorithm must construct a complete SLD-tree.

\subsection{Incomplete to Complete Information}
Many of the proposals in the literature on incomplete databases have
focussed on the extension of the relational model by the
introduction of null values. In this section, we show how view
update provides completion  of incomplete information. More detailed
surveys of this area can be found in \cite{Meyden}.

The earliest extension of the relational model to incomplete
information was that of Codd \cite{Codd} who suggested that missing
values should be represented in tables by placing a special
\emph{null value} symbol $'*'$  at any table location for which the
value is unknown. Table 3, shows an example of a database using this
convention. Codd proposed an extension to the relational algebra for
tables containing such nulls, based on three valued logic and a null
substitution principle.

In terms of our general semantic scheme, the intended semantics of a
database $D$ consisting of Codd tables can be described by defining
$Mod(D)$ to be the set of structures $M_{D'}$, where $D'$ ranges
over the relational databases obtained by replacing each occurrence
of $'*'$ in the database $D$ by some domain value. Different values
may be substituted for different occurrences.

A plausible integrity constraint on the meaning of a relational
operator on tables in $\mathcal{T}$ is that the result should be a
table that represents the set of relations obtained by pointwise
application of the operator on the models of these tables.  For
example, if $R$ and $S$ are tables in $\mathcal{T}$ then the result
of the join $R \Join S$ should be equal to a table T in
$\mathcal{T}$ such that

$$Mod(T)=\{r\Join t~|~r\in Mod(R),~ s\in Mod(S)\}$$

In case the definitions of the operators satisfy this integrity
constraint (with respect to the definition of the semantics \it Mod
\rm on $\mathcal{T}$).

Let us consider what above equation requires if we take $R$ and $S$
to be the Codd Tables 3.  First of all, note that in each model, if
we take the value of the null in the tuple (delhibabu,*) to be $v$,
then the join will contain one tuples (delhibabu, $v$), which
include the value $v$. If $T$ is to be a Codd table, it will need to
contain tuples (delhibabu,$X$) to generate each of these tuples,
where $X$ are either constants or '*'. We now face a problem. First,
$X$ cannot be a constant $c$, for whatever the choice of $c$ we can
find an instance $r\in Mod(R)$ and $s\in Mod(S)$ for which the tuple
(delhibabu, $c$) does not occur in $r\Join s$. If they were, $X$
would have their values in models of $T$ assigned independently.

Here the repetition of $*$ indicates that the \it same \rm value is
to be occurrence of the null in constructing a model of the table.
Unfortunately, this extension does not suffice to satisfy the
integrity constraint ($\forall x,y,z$ (y=x) $\leftarrow$
group\_chair(x,y) $\wedge$ group\_chair(x,z)).
\begin{center}
$\begin{array}{|c|c|}\hline
  \text{\rm Staff}&\text{\rm Group}\\\hline
  \text{\rm delhibabu}&\text{\rm infor1}\\
  \text{\rm delhibabu}&\text{\rm *}\\\hline
 \end{array}$
  $~~~~~~~~~~\begin{array}{|c|c|} \hline
  \text{\rm Group}&\text{\rm Chair}\\\hline
  \text{\rm infor1}&\text{\rm mattias}\\
  \text{\rm *}&\text{\rm aravindan}\\  \hline
 \end{array}$
\end{center}

\begin{center} \text{\bf Tab. 3. \rm Base Table after Transaction}\end{center}

In the model of these tables in which $*=infor1$,  the join contains
the tuple (delhibabu, infor1) and (infor1, aravindan).

$$\text{\rm If}~*_1=\text{\rm infor1 then (delhibabu, infor1)}\in R\Join S$$
$$\text{\rm If}~*_2=\text{\rm infor1 then (infor1, aravindan)}\in R\Join S$$

The following table shows when transaction is made to base table:

$$\begin{array}{|c|c|c|}\hline
\text{\rm Staff}&\text{\rm Group}&\text{\rm Chair}\\\hline
  \text{\rm delhibabu}& \text{\rm infor1} &\text{\rm mattias}\\
\text{\rm delhibabu}& \text{\rm * } &\text{\rm aravindan}\\\hline
\end{array}$$

\begin{center} \text{\bf Tab. 4. \rm $s\otimes r$  after Transaction }\end{center}

The following table shows completion of incomplete information with
application of integrity constraint and redundancy:

$$\begin{array}{|c|c|c|}\hline
\text{\rm Staff}&\text{\rm Group}&\text{\rm Chair}\\\hline
  \text{\rm delhibabu}& \text{\rm infor1} &\text{\rm aravindan}\\\hline
\end{array}$$
\begin{center} \text{\bf Tab. 5.\rm~Redundant Table}\end{center}

%%%--------------------------------------------------------------------------------

\section{Related Works}

We begin by recalling previous work on view deletion. Chandrabose
\cite{Arav1,Arav}, defines a contraction operator in view deletion
with respect to a set of formulae or sentences using Hansson's
\cite{Hans2} belief change. Similar to our approach, he focused on
set of formulae or sentences in knowledge base revision for view
update wrt. insertion and deletion and formulae are considered at
the same level. Chandrabose proposed different ways to change
knowledge base via only database deletion, devising particular
postulate which is shown to be necessary and sufficient for such an
update process.

Our Horn knowledge base consists of two parts, immutable part and
updatable part , but focus is on principle of minimal change. There
are more related works on that topic. Eiter \cite{Eit},
Langlois\cite{Lang}, and Delgrande \cite{Delg} are focusing on Horn
revision with different perspectives like prime implication, logical
closure and belief level. Segerberg \cite{Seg} defined new modeling
for belief revision in terms of irrevocability on prioritized
revision. Hansson \cite{Hans2}, constructed five types of
non-prioritized belief revision. Makinson \cite{Mak} developed
dialogue form of revision AGM. Papini\cite{Pap} defined a new
version of knowledge base revision. Here, we consider immutable part
as a Horn clause and updatable part as an atom(literals).

We are bridging gap between philosophical work, paying little
attention to computational aspects of database work. In such a case,
Hansson's\cite{Hans2} kernel change is related with abductive
method. Aliseda's \cite{Alis} book on abductive reasoning is one of
the motivation keys. Christiansen's \cite{Chris} work on dynamics of
abductive logic grammars exactly fits our minimal change (insertion
and deletion). Wrobel's \cite{Wrob} definition of  first order
theory revision was helpful to frame our algorithm.

On other hand, we are dealing with view update problem. Keller's
\cite{Kell} thesis is motivation  for view update problem. There is
a lot of papers on view update problem (for example, recent survey
paper on view update by Chen and Liao\cite{Chen}  and survey paper
on view algorithm by Mayol and Teniente \cite{Mayol}. More similar
to our work is paper presented by Bessant et al. \cite{Bess} , local
search-based heuristic technique that empirically proves to be often
viable, even in the context of very large propositional
applications. Laurent et al.\cite{Lau}, parented updating deductive
databases in which every insertion or deletion of a fact can be
performed in a deterministic way.

Furthermore, and at a first sight more related to our work, some
work has been done on ontology systems and description logics (Qi
and Yang \cite{Qi}, and Kogalovsky \cite{Kog}). Finally, when we
presented connection between belief update versus database update,
we did not talk about complexity (see the works of Liberatore
\cite{Lib1,Lib2}, Caroprese \cite{Caro}, Calvanese's \cite{Cal}, and
Cong \cite{Cong}).

The significance of our work can be summarized in the following:
\begin{description}
  \item[-] We have defined new kind of revision operator on
  knowledge base and obtained axiomatic characterization for it.
  This operator of change is based on $\alpha$ consistent-remainder set.
  Thus, we have presented a way to construct revision operator
  without need to make use of the generalized Levi's identity nor
  of a previously defined contraction operator.
  \item[-] We have defined new way of insertion and deletion of an atom(literals) as per
  norm of principle of minimal change.
  \item[-] We have proposed new generalized revision algorithm for
  knowledge base dynamics, interesting connections with kernel change and
  abduction procedure.
   \item[-] We have written new view insertion algorithm for DDB, and
  we provided Horn knowledge base revision, using our axiomatic method.
   \item[-] Finally, we shown connection between belief update versus
  database update.
\end{description}

%%%%%------------------------------------------------------------------------------------------

\section{Conclusion and remarks}

The main contribution of this research is to provide a link between
theory of belief dynamics and concrete applications such as view
updates in databases. We argued for generalization of belief
dynamics theory in two respects: to handle certain part of knowledge
as immutable; and dropping the requirement that belief state be
deductively closed. The intended generalization was achieved by
introducing the concept of knowledge base dynamics and generalized
contraction for the same. Further, we also studied the relationship
between  knowledge base dynamics and abduction resulting in a
generalized algorithm for revision based on abductive procedures. We
also successfully demonstrated how knowledge base dynamics can
provide an axiomatic characterization for insertion an
atom(literals) to a definite deductive database. Finally, we give a
quick overview of the main operators for belief change, in
particular, belief update versus database update.

In bridging the gap between belief dynamics and view updates, we
have observed that a balance has to be achieved between
computational efficiency and rationality. While rationally
attractive notions of generalized revision prove to be
computationally inefficient, the rationality behind efficient
algorithms based on incomplete trees is not clear at all. From the
belief dynamics point of view, we may have to sacrifice some
postulates, vacuity for example, to gain computational efficiency.
Further weakening of relevance has to be explored, to provide
declarative semantics for algorithms based on incomplete trees.

On the other hand, from the database side, we should explore various
ways of optimizing the algorithms that would comply with the
proposed declarative semantics. We believe that partial deduction
and loop detection techniques, will play an important role in
optimizing algorithms of the previous section. Note that, loop
detection could be carried out during partial deduction, and
complete SLD-trees can be effectively constructed wrt a partial
deduction (with loop check) of a database, rather than wrt database
itself. Moreover, we would anyway need a partial deduction for
optimization of query evaluation.

Though we have discussed only about view updates, we believe that
knowledge base dynamics can also be applied to other applications
such as view maintenance, diagnosis, and we plan to explore it
further (see works \cite{Caro} and \cite{Bis}). It would also be
interesting to study how results using soft stratification
\cite{Beh} with belief dynamics, especially the relational approach,
could be applied in real world problems. Still, a lot of
developments are possible, for improving existing operators or for
defining new classes of change operators. As immediate extension,
question raises: is there any \emph{real life application for AGM in
25 year theory?} \cite{Ferme}. The revision and update are more
challenging in logical view update problem(database theory), so we
can extend the theory to combine results similar to Hansson's
\cite{Hans1}, Konieczny's \cite{Kon}and Nayak, \cite{Nayak2}.

%%---------------------------------------------------------------------------------------------------

\section*{Appendix}

\text{Proof of Theorem 1}. (\textbf{If part})~* satisfies (KB*1) to
(KB*6) and (KB*7.3). We must show that $*$ is a generalized kernel
revision. Let $\sigma$ be a incision function such that for
$\alpha$. When $KB_{I}\vdash\alpha$, (KB*1) to (KB*6) and (KB*7.3)
imply that $KB*\alpha=KB$ coincides with generalized revision and
follow PMC.

When $KB_{I}\vdash\neg\alpha$, the required result follows from the
two observations:
\begin{itemize}
\item[1.] $\exists KB'\in KB\bot_{\bot}\alpha$ s.t.$KB*\alpha\subseteq
KB'$ (when $KB_{I}\vdash \alpha$)\\
Let $\sigma$ be an incision function for $KB$ and $*_\sigma$ be the
generalized revision on $KB$ that is generated by $\sigma$. Since
* satisfies closure (KB*1), $KB*_{\sigma}\alpha$ is KB contained in $\alpha$.
Also, satisfaction of weak success postulate (KB*2) ensures that
$\alpha\subseteq KB*_{\sigma}\alpha$. Every element of
$KB\bot_{\bot}\alpha$ is a inclusion minimal subset that does derive
$\alpha$, and so any subset of KB that does derive $\alpha$ must be
contained in a member of $KB\bot_{\bot}\alpha$.

\item[2.] $\bigcap(KB\bot_{\bot}\alpha)\subseteq KB*_{\sigma}\alpha$
(when $KB_{I}\vdash \alpha$)\\
Consider any $\beta \in \bigcap(KB\bot_{\bot}\alpha)$. Assume that
$\beta\not\in KB*\alpha$. Since * satisfies weak relevance postulate
(KB*7.3), it follows that there exists a set KB' s.t. $KB'\subseteq
KB\cup \alpha$; $KB'$ is a consistent with $\alpha$; and $KB' \cup
\{\beta\}$ is inconsistent with $\alpha$. But this contradicts that
$\beta$ is present in every minimal subset of KB that does derive
$\alpha$. Hence $\beta$ must not be in $KB*_{\sigma}\alpha$.
\end{itemize}

(\textbf{Only if part})~Let $KB*\alpha$ be a generalized revision of
$\alpha$ for KB. We have to show that $KB*\alpha$ satisfies the
postulate (KB*1) to (KB*6) and (KB*7.3).

Let $\sigma$ be an incision function for $KB$ and $*_\sigma$ be the
generalized revision on $KB$ that is generated by $\sigma$.

\begin{description}
\item[Closure] Since $KB*_{\sigma}\alpha$ is a Horn knowledge base, this
postulate is trivially shown.

\item[Weak Success] Suppose that $\alpha$ is consistent. Then it is trivial by definition that $\alpha \subseteq
KB*_{\sigma} \alpha$.

\item[Inclusion] Trivial by definition.

\item[Immutable-inclusion] Since every $X \in KB\bot_{\bot}
\alpha$ is such that $X \subseteq KB_I$ then this postulate is
trivially shown.

\item[Vacuity 1] Trivial by definition.

\item[Vacuity 2] If $KB\cup \{\alpha\}$ is consistent then
$KB\bot_{\bot} \alpha$ = \{\{KB\}\}. Hence $KB*_{\sigma}\alpha$ =
$KB\cup \{\alpha\}$.

\item[Consistency] Suppose that $\alpha$ is consistent. Then $KB \bot_{\bot} \alpha \neq= \emptyset$ and by
definition, every $X \in KB \bot_{\bot} \alpha$ is consistent with
$\alpha$. Therefore, the intersection of any subset of $KB
\bot_{\bot} \alpha$ is consistent with $\alpha$. Finally,
$KB*_{\sigma} \alpha$ is consistent.

\item[Uniformity] If $\alpha$ and $\beta$ are KB-equivalent, then $KB\bot_{\bot}
\alpha$ = $KB\bot_{\bot} \beta$

\item[Weak relevance] Let $\beta\in KB$ and $\beta\notin KB*_{\sigma}\alpha$. Then
$KB*_{\sigma}\alpha \neq KB$ and, from the definition of
$*_{\sigma}$,it follows that:\\

\hspace{2.5cm}$KB*_{\sigma}\alpha$=$(KB\backslash\sigma(KB\bot_{\bot}\alpha))\cup\alpha$\\

Therefore, from $\beta\not\in
(KB\backslash\sigma(KB\bot_{\bot}\alpha))\cup\alpha$ and $\beta\in
KB$, we can conclude that $\beta\in\sigma(KB\bot_{\bot}\alpha)$. By
definition $\sigma(KB\bot_{\bot}\alpha) \subseteq \bigcup
KB\bot_{\bot}\alpha$, and it follows that there is some $X\in
KB\bot_{\bot}\alpha$ such that $\beta \in X$. X is a minimal
KB-subset inconsistent with $\alpha$. Let $Y=X\backslash \{\beta\}$.
Then Y is such that $Y\subset X\subseteq KB \subseteq KB\cup\alpha$.
Y is consistent with $\alpha$ but $Y\cup\{\beta\}$ is consistent
with $\alpha$. $\blacksquare$
\end{description}

\vspace{0.1cm}

\text{Proof of Theorem 2}. Follows from Theorem \ref{T7} and
Definition \ref{D21}. $\blacksquare$

\vspace{0.1cm}

\text{Proof of Lemma 1}.
\begin{enumerate}
\item[1.] Consider a $\Delta (\Delta\in\Delta_i \cup \Delta_j)\in S$.
 We need to show that $\Delta$ is generated by algorithm
3 at step 2. From lemma 1, it is clear that there exists a
$A$-kernel $X$ of $DDB_G$ s.t. $X \cap EDB = \Delta_j$ and $X \cup
EDB = \Delta_i$. Since $X \vdash A$, there must exist a successful
derivation for $A$ using only the elements of $X$ as input clauses
and similarly $X \nvdash A$. Consequently $\Delta   $ must have been
constructed at step 2.
\item[2.] Consider a $\Delta'((\Delta'\in\Delta_i \cup \Delta_j)\in S'$. Let $\Delta'$ be
constructed from a successful(unsuccessful) branch $i$ via
$\Delta_i$($\Delta_j$). Let $X$ be the set of all input clauses used
in the refutation $i$. Clearly $X\vdash A$($X\nvdash A$). Further,
there exists a minimal (wrt set-inclusion) subset $Y$ of $X$ that
derives $A$ (i.e. no proper subset of $Y$ derives $A$). Let $\Delta
= Y \cap EDB$ ($Y \cup EDB$). Since IDB does not(does) have any unit
clauses, $Y$ must contain some EDB facts, and so $\Delta$ is not
empty (empty) and obviously $\Delta\subseteq \Delta'$. But, $Y$ need
not (need) be a $A$-kernel for $IDB_G$ since $Y$ is not ground in
general. But it stands for several $A$-kernels with the same
(different) EDB facts $\Delta$ in them. Thus, from lemma 1, $\Delta$
is a DDB-closed locally minimal abductive explanation for $A$ wrt
$IDB_G$ and is contained in $\Delta'$.
\item[3.] Since this proof requires some details of acyclic programs
that are not directly related to our discussion here, it is
relegated [9].
\end{enumerate}

\vspace{0.1cm}

\text{Proof of Lemma 2}.
\begin{enumerate}
\item[1.] (\textbf{Only if part})~Suppose $H$ is a minimal hitting set for $S$. Since $S
\subseteq S'$ , it follows that $H \subseteq \bigcup S'$ . Further,
$H$ hits every element of $S'$ , which is evident from the fact that
every element of $S'$ contains an element of $S$. Hence $H$ is a
hitting set for $S'$ . By the same arguments, it is not difficult to
see that $H$ is minimal for $S'$ too.\\

(\textbf{If part})~Given that $H$ is a minimal hitting set for $S'$
, we have to show that it is a minimal hitting set for $S$ too.
Assume that there is an element $E \in H$ that is not in $\bigcup
S$. This means that $E$ is selected from some $Y \in S'\backslash
S$. But $Y$ contains an element of $S$, say $X$. Since $X$ is also a
member of $S'$ , one member of $X$ must appear in $H$. This implies
that two elements have been selected from $Y$ and hence $H$ is not
minimal. This is a contradiction and hence $H \subseteq \bigcup S$.
Since $S \subseteq S'$ , it is clear that $H$ hits every element in
$S$, and so $H$ is a hitting set for $S$. It remains to be shown
that $H$ is minimal. Assume the contrary, that a proper subset $H'$
of $H$ is a hitting set for $S$. Then from the proof of the only if
part, it follows that $H'$ is a hitting set for $S'$ too, and
contradicts the fact that $H$ is a minimal hitting set for $S'$ .
Hence, $H$ must be a minimal hitting set for $S$.\\

\item[2.] (\textbf{If part})~Given that $H$ is a hitting set for $S'$ , we have to
show that it is a hitting set for $S$ too. First of all, observe
that $\bigcup S = \bigcup S'$ , and so $H \subseteq \bigcup S$.
Moreover, by definition, for every non-empty member $X$ of $S'$ , $H
\cap X$ is not empty. Since $S \subseteq S'$ , it follows that $H$
is a hitting set for $S$ too.\\

(\textbf{Only if part})~Suppose $H$ is a hitting set for $S$. As
observed above, $H \subseteq \bigcup S'$ . By definition, for every
non-empty member $X\in S$, $X \cap H$ is not empty. Since every
member of $S'$ contains a member of $S$, it is clear that $H$ hits
every member of $S'$ , and hence a hitting set for $S'$ .
$\blacksquare$
\end{enumerate}

\vspace{0.1cm}

\text{Proof of Theorem 3}. From Lemma \ref{l2}, it is clear that
step 1 and 2 generate all DDB-closed locally minimum abductive
explanation for $\mathcal{A}$ wrt $IDB_G$ and some additional sets
that contain a DDB-closed local minimal abductive explanation for
$\mathcal{A}$ wrt $IDB_G$. Step 3 and 4 clearly computes an
inclusion-minimal hitting set for this. This required now following
Lemma \ref{l3} and Theorem \ref{T8}. $\blacksquare$

\vspace{0.1cm}

\text{Proof of Theorem 4}. Follows from Lemma \ref{l2}, Lemma
\ref{l3}, and Theorem \ref{T8}. $\blacksquare$

%%%---------------------------------------------------------------------------------------------------------------------

%%%%-------------------------------------------------------------------------------------------------------------------

\end{document}